\documentclass[twocolumn,showpacs,showkeys,superscriptaddress]{revtex4}

\usepackage{amsmath}
\usepackage{bbold}
\usepackage{amsfonts}
\usepackage{amssymb}
\usepackage{pbsi}
\usepackage[T1]{fontenc}
\usepackage{hyperref}
\usepackage{xcolor}
\usepackage{graphicx}
\usepackage{subfig}
\usepackage{float}

\def\openone{\leavevmode\hbox{\small1\kern-3.8pt\normalsize1}}
\def\N{\leavevmode\hbox{ Z \kern-8 pt\normalsize{Z}}}
\def\openone{\leavevmode\hbox{\small1\kern-3.8pt\normalsize1}}
\def\openJ{\leavevmode\hbox{J \kern-9.5pt\normalsize J}}
\def\openS{\leavevmode\hbox{ S \kern-9.3pt\normalsize S}}

\newcommand{\bb}{\begin{equation}}
\newcommand{\ee}{\end{equation}}
\newcommand{\eqb}{\begin{eqnarray}}
\newcommand{\eqf}{\end{eqnarray}}

\usepackage{color}

\begin{document}

\title{Heat bullets}

\author{Felipe A. Asenjo}
\email{felipe.asenjo@uai.cl}
\affiliation{Facultad de Ingenier\'ia y Ciencias,
Universidad Adolfo Ib\'a\~nez, Santiago 7491169, Chile.}
\author{Sergio A. Hojman}
\email{sergio.hojman@uai.cl}
\affiliation{Departamento de Ciencias, Facultad de Artes Liberales,
Universidad Adolfo Ib\'a\~nez, Santiago 7491169, Chile.}
\affiliation{Departamento de F\'{\i}sica, Facultad de Ciencias, Universidad de Chile,
Santiago 7800003, Chile.}
\affiliation{Centro de Recursos Educativos Avanzados,
CREA, Santiago 7500018, Chile.}

\begin{abstract}
New localized structured solutions for the three-dimensional linear diffusion (heat) equation are presented. These new solutions are written in terms of Airy functions and either Gaussian or Bessel functions. They accelerate along their propagation direction, while in the plane orthogonal to it, they retain their either Gaussian or Bessel structure. These diffusion (heat) densities retain a localized structure in space as they propagate,  and may be considered the heat analogue of Airy light  bullets. 
\end{abstract}


\maketitle

\section{Introduction}

In 1979, Berry and Balazs found free particle  accelerating solutions to the one-dimensional Schrödinger equation \cite{berry}. This solution has the same global properties
as usual plane wave solutions for free particle \cite{lekner}. However, it shows  drastic local differences, that displays the  accelerating properties of a propagating particle.
This unexpected and surprising prediction 
has been studied and confirmed for electrons \cite{bloch,harris,OriReinhardt,JianXing,idokanemi,Goutsoulas,Grillo}, and it has been also proved to be a fundamental characteristic for light  \cite{sivi,irivka,Efremidis,hacyan,Aleahmad,Chremmos,Kaminer,Bandresm,Patsyk11,mabrandes,EsatKondakci,Baumgartl},
plasma \cite{lWu,ychen,Polynkin1, Panagiotopoulos,Gorbunov}, acoustic \cite{zlin1,UriBarZiv,xsli,zhaoh,Kawanaka} 
and gravitational \cite{asenjHojmanGravWave} propagation, among others.

In the same spirit, it has been recently shown that there exist accelerating solutions to the linear diffusion (heat) equation \cite{asenjohojmanT}. This fact seems to imply that the existence of accelerating solutions is widely present in {\it{parabolic linear}} equations describing {\it{free}} wave evolution, as it has already been confirmed to occur in physical systems involving electrons, light, plasmas or gravitation, as mentioned above. The purpose of this work is to study a new kind of accelerating phenomenon in the linear diffusion equation that mimics light bullets.

One of the most interesting constructions that can be achieved for a propagating field with accelerating Airy properties are the {\it bullet} solutions. These are localized structures that propagate in a non-dispersive accelerated fashion. 
These Airy bullets have been theoretically described and experimentally confirmed for light  \cite{GeorgiosASiviloglou,YaronSilberberg,achong,Daryoush,Peeter,ypeng,Zhenkun,xipengetal,Christodoulides222,WeiPingZhong}. 
In this Letter we show that the diffusion equation admits diffusive (heat) bullet solutions which are the direct analogue of light bullets. This is a new form of wave-like propagation for diffusion, complementing other forms of heat wave transport \cite{barcelona,JaeHyukChoi1}. 
In this way, these solutions can be understood as heat bullets with diffusive features that are inherited from the diffusive medium where they propagate. Contrary to light bullets features, diffusive behavior cannot be avoided in these kinds of accelerated solutions. They can have diffusive properties in the three-dimensional space or along the direction of propagation only.

The goal is to find diffusive bullet solutions for the three-dimensional diffusion equation 
\begin{equation}
    \frac{\partial \phi}{\partial t}=D_1 \frac{\partial^2 \phi}{\partial x^2}+D_2 \frac{\partial^2 \phi}{\partial y^2}+D_3 \frac{\partial^2 \phi}{\partial z^2}\, ,
    \label{ecdifusion}
\end{equation}
for a diffusing density $\phi(t,{\bf r})$. Here, the constant diffusion coefficients in each of the cartesian axes $D_i$ ($i=1,2,3$) are arbitrary. 
Below, we will show that is possible to obtain accelerating propagating solutions of $\phi(t,{\bf r})$  that have localized density structures in space that do not propagate as standard diffusion. What we look for are solutions travelling in a preferred direction (say in the  $z$-direction) in an accelerated fashion, while in the transverse $(x-y)$ plane it remains structured.

\section{Airy-Gauss diffusive bullet}

We look for solutions that remain structured in the transverse $(x-y)$ plane in the usual diffusive Gaussian-like form.  
Under this assumption, the solution for density propagates in an accelerated fashion along a longitudinal direction in the form of an Airy function. 

Let us assume that the diffusion density has the form 
$\phi(t,x,y,z)=\varphi(t,z)\psi(t,x,y)$, such that $\psi$ fulfill
$\partial_t \psi=D_1 \partial^2_x \psi+D_2 {\partial^2_y \psi}$, allowing a diffusive Gaussian solution in the transverse plane. 
In this way, $\varphi$ now satisfies the equation $\partial_t \varphi=D_3 \partial^2_z \varphi$. In Ref.~\cite{asenjohojmanT}, it has been shown that this equation has accelerating solutions in form of Airy  
    functions ${\mbox{Ai}}$.
Therefore, the complete solution of Eq.~\eqref{ecdifusion} for the diffusion density becomes
\begin{eqnarray}
&& \phi(t,x,y,z)=\frac{\phi_0}{\sqrt{D_1D_2} \, t}\, {\mbox{Ai}}\left(k\frac{z}{\sqrt{D_3}}+k^4 t^2\right)\nonumber\\
    &&\quad\times\exp\left(k^3t\frac{z}{\sqrt{D_3}}+\frac{2}{3}k^6 t^3 -\frac{x^2}{4D_1 t}-\frac{y^2}{4D_2 t} \right)\, ,
    \label{Airygauss}
\end{eqnarray}
where $\phi_0$ is a constant, and
$k$ is an arbitrary constant with units of inverse of square root of time. This density can be proved to be solution by directly inserting it in Eq.~\eqref{ecdifusion}. 

\begin{figure}[ht]
  \includegraphics[height=75mm]{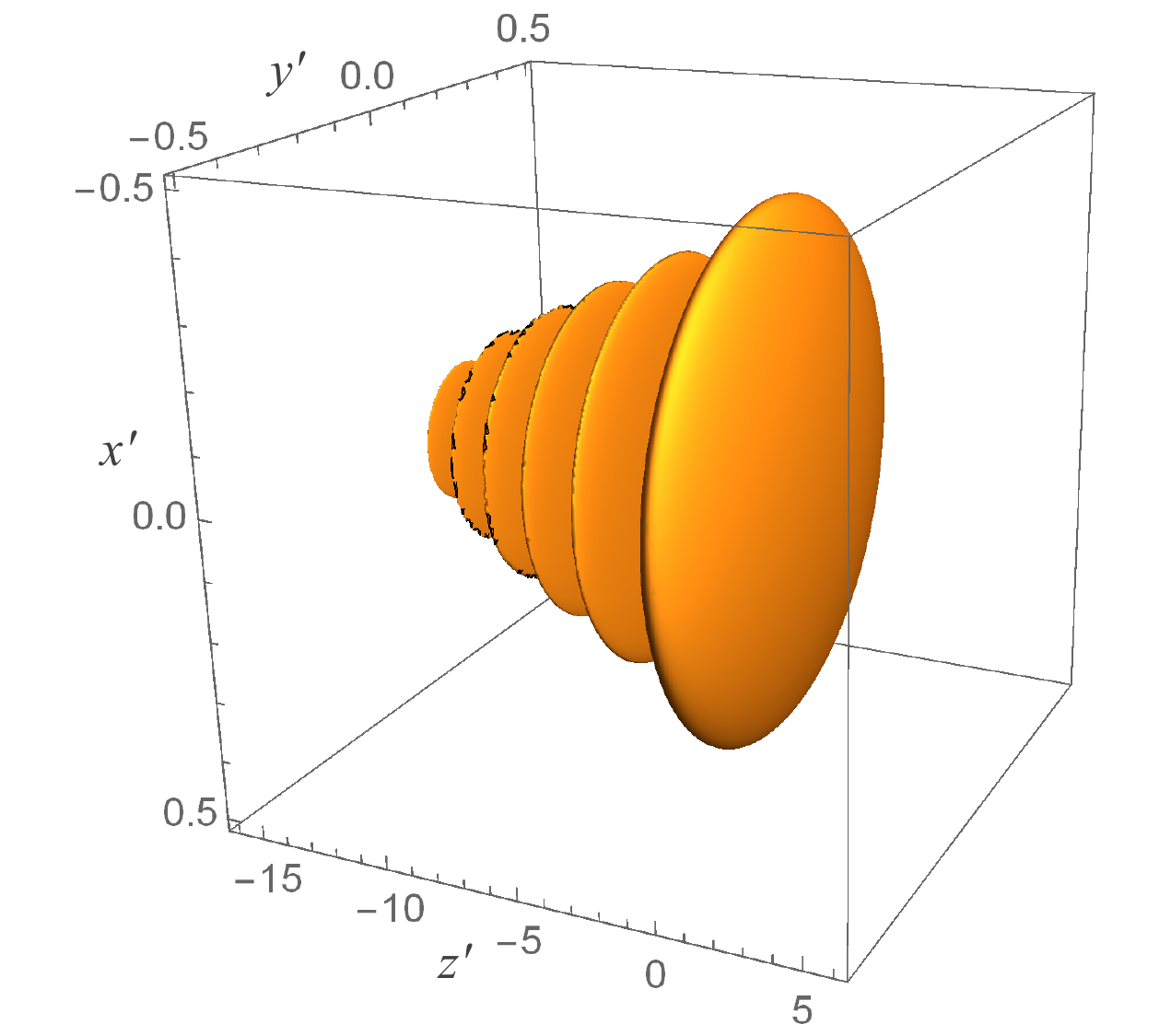} \\
  \includegraphics[height=75mm]{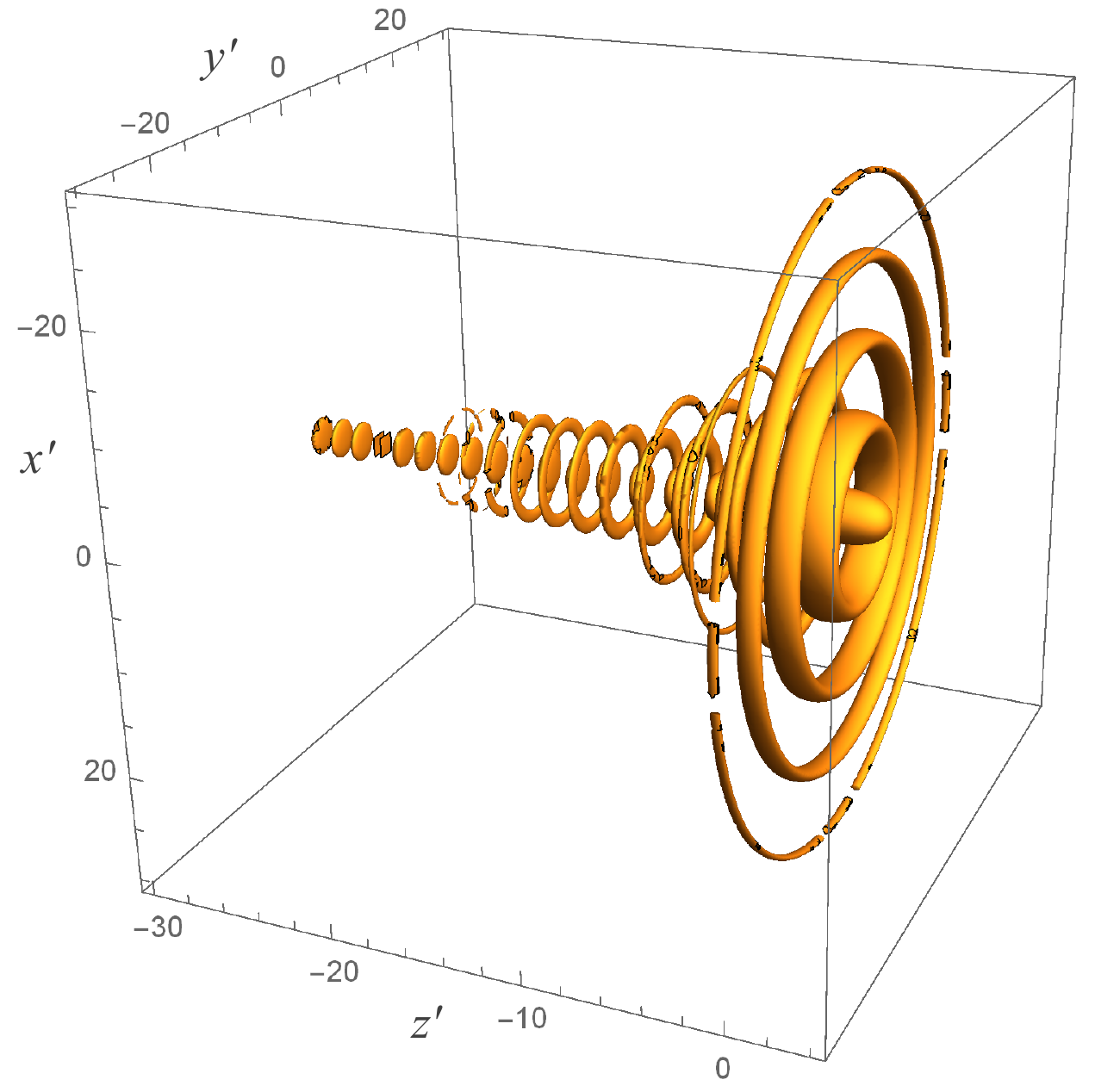}
  \caption{(a) Iso-contour plot for  density \eqref{Airygauss}, such that $\phi \sqrt{D_1 D_2}/(\phi_0 k^2)=4$, in terms of variables  $x'=k\, x/\sqrt{D_1}$, $y'=k\, y/\sqrt{D_2}$ and $z'=k\, z/\sqrt{D_3}$, for $k^2 t=0.04$. (b) Iso-contour plot for  density \eqref{soluDensiAiryBess}, such that $\phi/\phi_0=0.08$, in terms of variables $x'$, $y'$ and $z'$, for $k^2 t=0.01$ and $\lambda/k^2=0.8$. 
}
\label{fig1}. 
\end{figure}

Diffusion density \eqref{Airygauss} propagates as an accelerating localized structure. Its form is analogue to the known Airy light bullets, and may, therefore, be considered as  diffusing heat bullets. This is seen in Fig.~\ref{fig1}(a), where  an iso-contour plot for density \eqref{Airygauss} is displayed for a given time $k^2 t=0.04$.
Different maxima and minima of this  bullet
show acceleration along the longitudinal direction, due to the argument on the Airy function.
This acceleration can be better seen in its maximum density lobe, as we will see below in Sec.~\ref{accesection}. 

The evolution of  solution \eqref{Airygauss} is shown as density plots in Figs.~\ref{fig2}(a), (b) and (c), where this density   is shown in the $x=0$ plane  for three different times. As time elapses, the solution is described by a unique maximum density lobe. This is a characteristic of Airy functions. We can see that the position of the maximum density lobe changes in time (depicted by red vertical lines). This is analyzed in Sec.~\ref{accesection} as well.

\section{Airy-Bessel diffusive bullet}

A different localized structured solution can be constructed from the same linear diffusion equation \eqref{ecdifusion}.

The transverse part of the diffusive density may be modelled by a Bessel function. We   
 assume that the diffusion density has the form
$\phi(t,x,y,z)=\varphi(t,z)\psi(x,y)$, requiring that $\psi$ fulfill
$D_1 \partial^2_x \psi+D_2 {\partial^2_y \psi}+\lambda\psi=0$, where $\lambda$ is an arbitrary constant with  units of  inverse of time. This implies that this diffusive solution behaves as a Bessel function 
of zeroth order $J_0$
 in the transverse plane. 
Also, this implies that
 $\varphi$ now satisfies the equation $\partial_t \varphi=D_3 \partial^2_z \varphi-\lambda\varphi$, which again can be solved by an Airy function with accelerating properties in $z$-direction \cite{asenjohojmanT}.

Thereby, the above form for the solution allows us
to find
 a diffusion density  with the form
\begin{eqnarray}
\phi(t,x,y,z)&=&\phi_0\, {\mbox{Ai}}\left(k\frac{z}{\sqrt{D_3}}+k^4 t^2\right)\nonumber\\
    &&\times \exp\left(k^3t\frac{z}{\sqrt{D_3}}-\lambda\,  t+\frac{2}{3}k^6 t^3 \right)\nonumber\\
    &&\times J_0\left(\sqrt{\frac{\lambda\, x^2}{D_1}+\frac{\lambda\,  y^2}{D_2}} \right)\, ,
    \label{soluDensiAiryBess}
\end{eqnarray}
where $\phi_0$ is a constant, and
$k$ is anew an arbitrary constant with  units of inverse of square root of time.

\begin{figure*}[ht]
\begin{tabular}{ccc}
  \includegraphics[height=45mm]{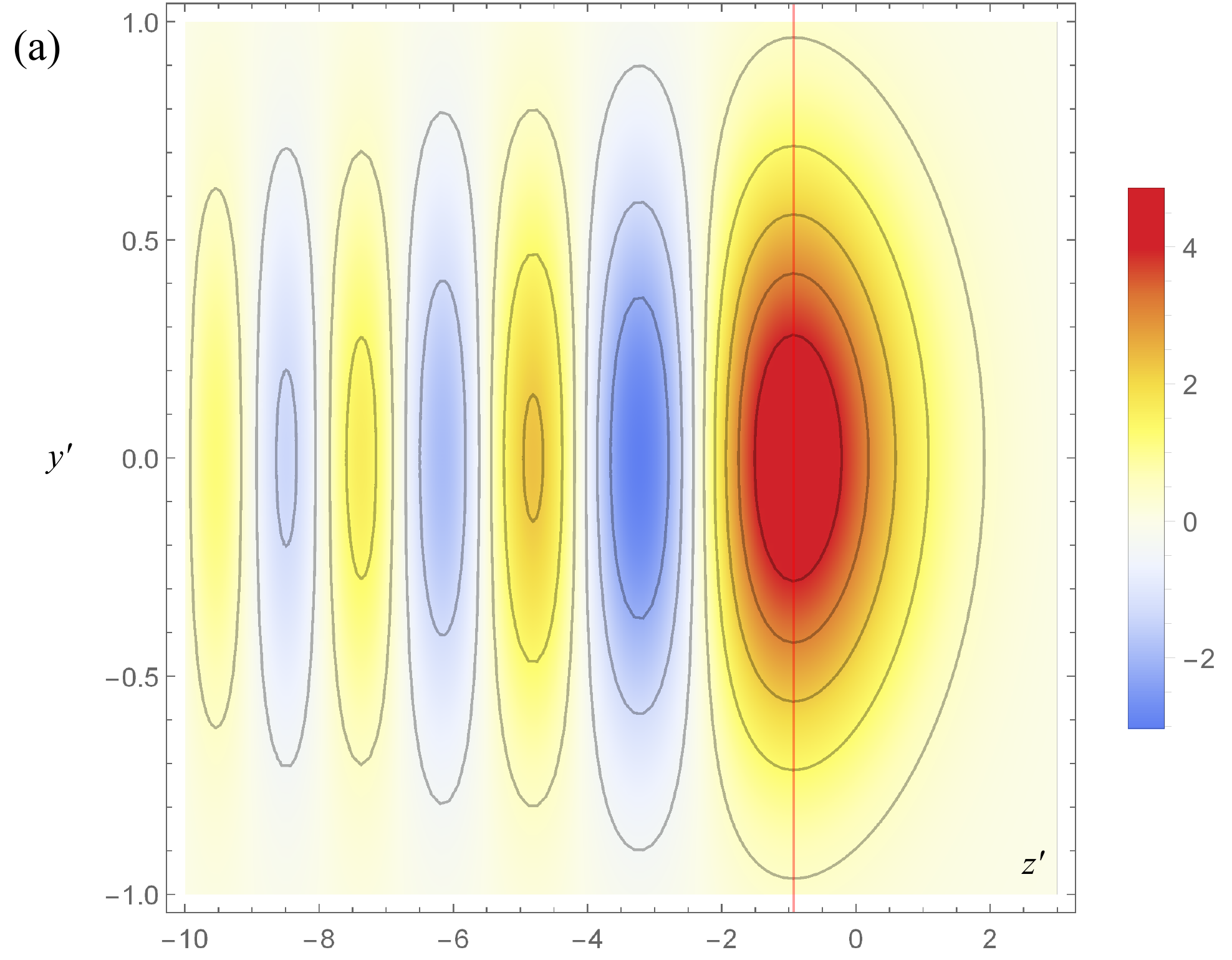} &     \includegraphics[height=45mm]{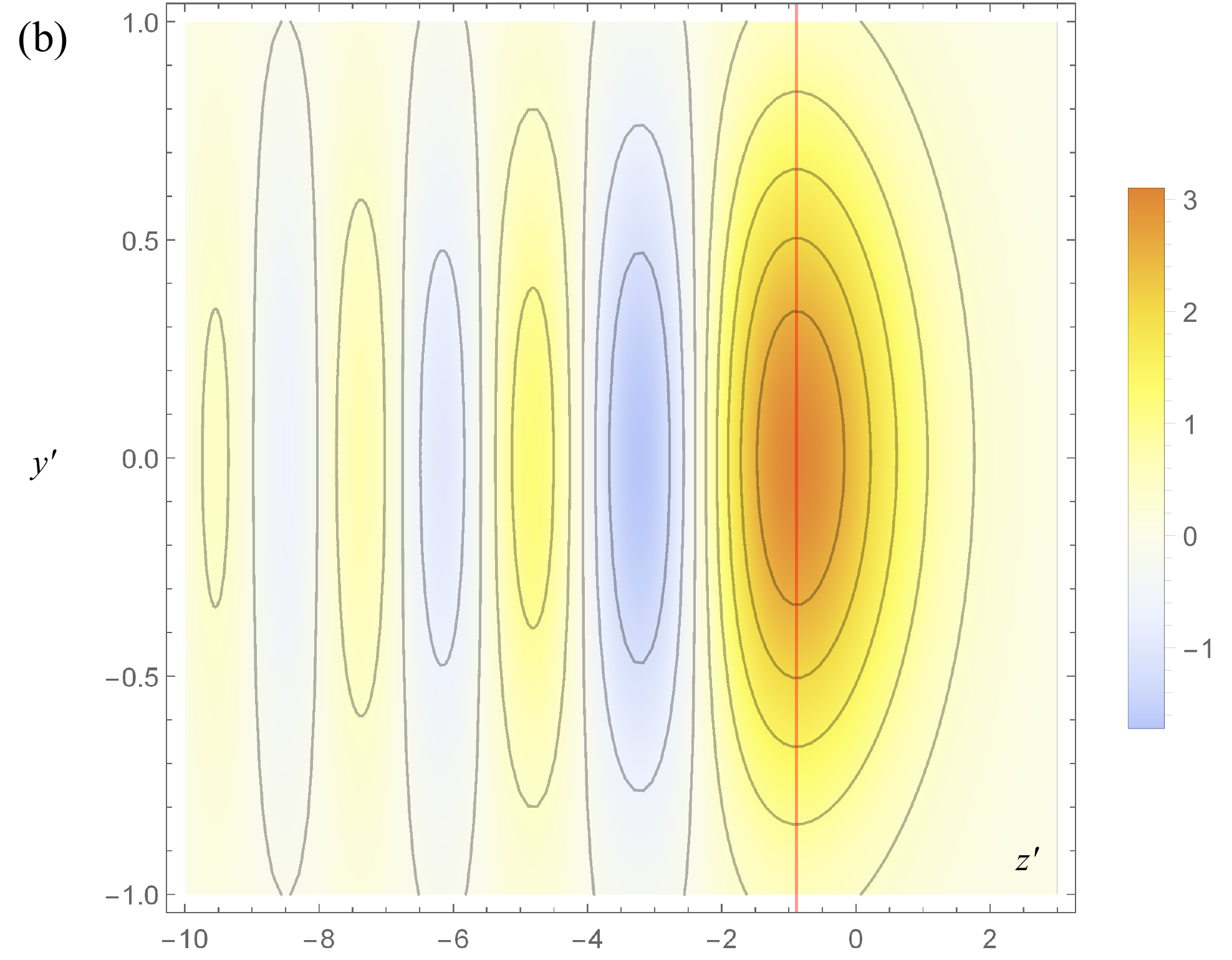} &\includegraphics[height=45mm]{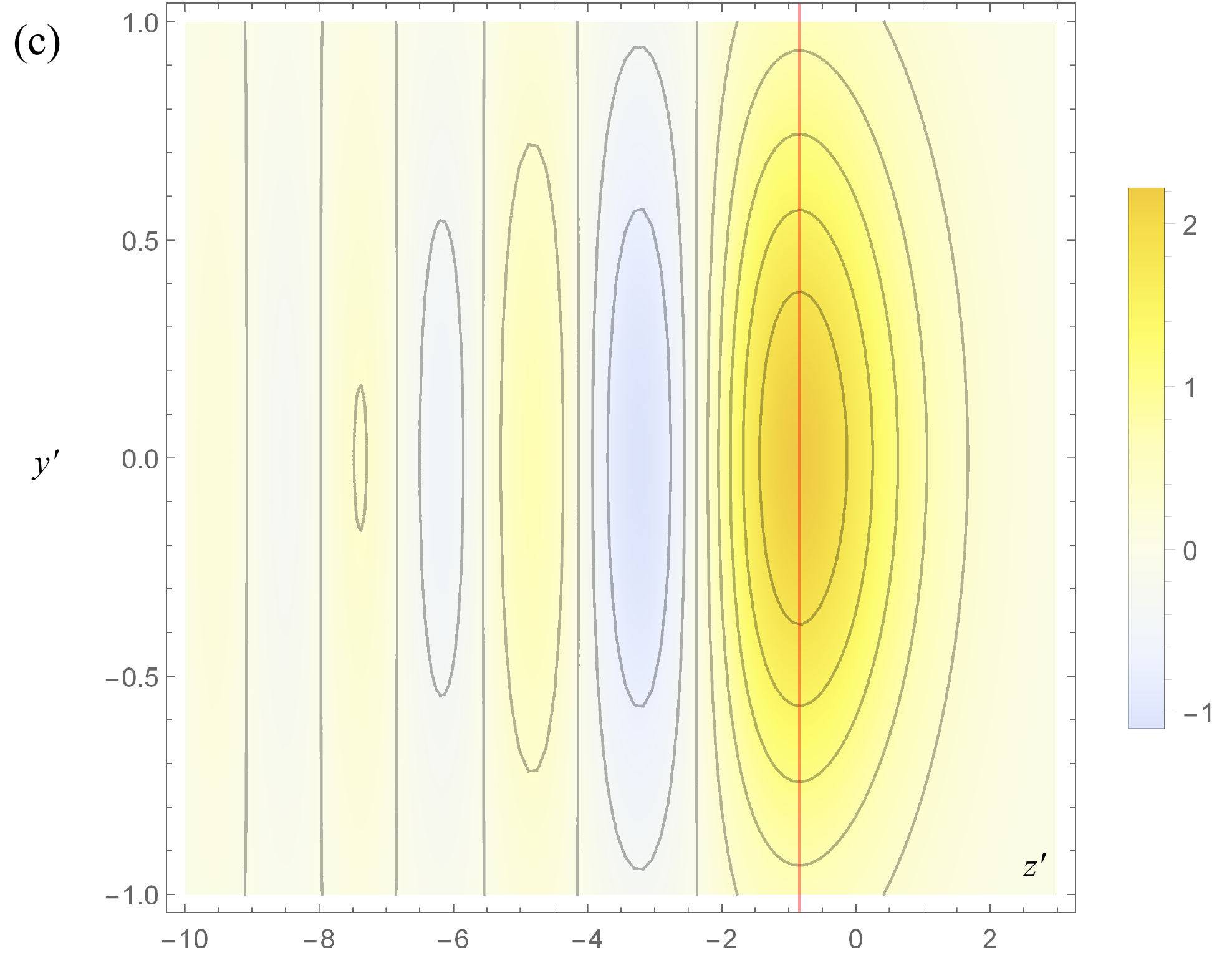} \\
 \includegraphics[height=45mm]{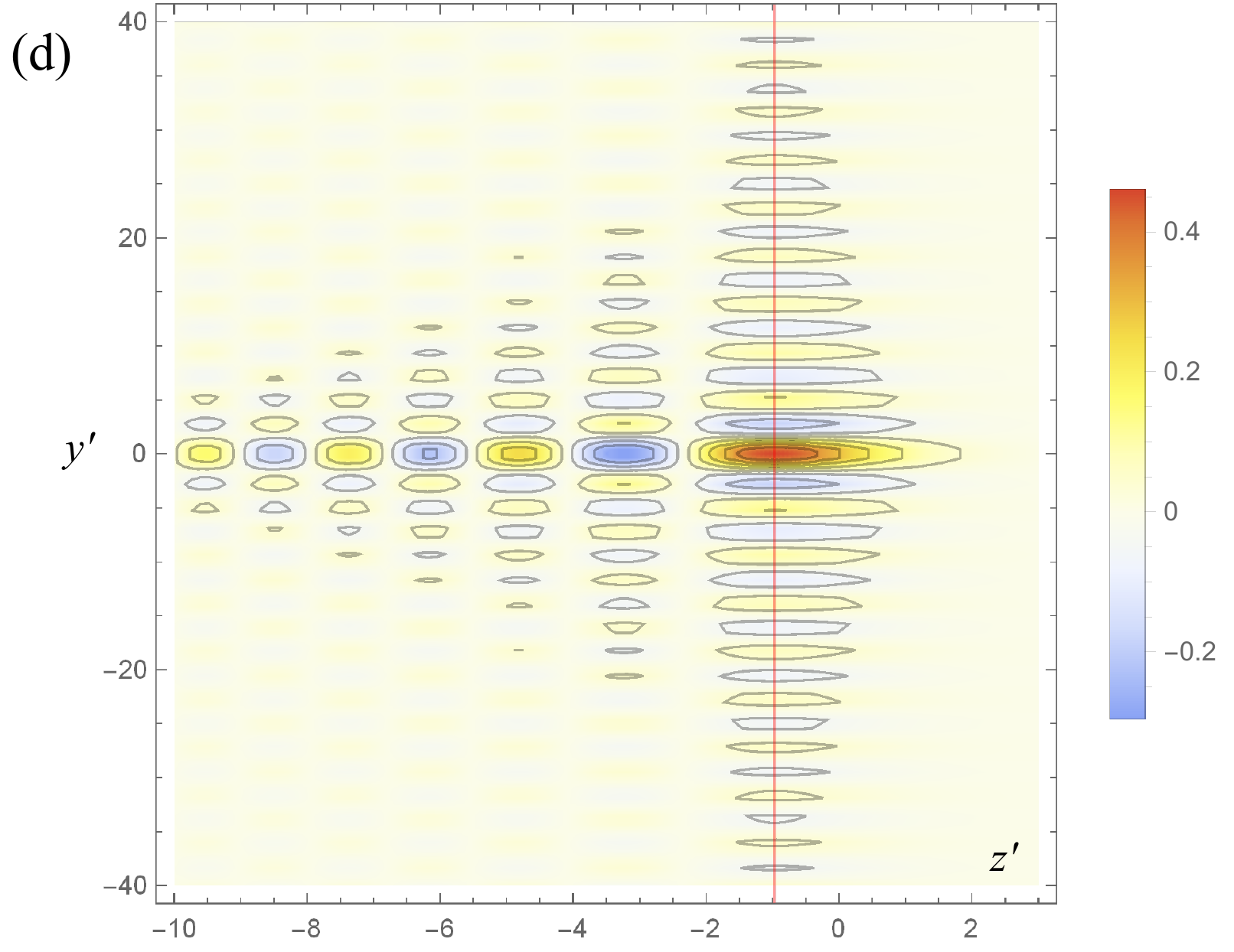} & \includegraphics[height=45mm]{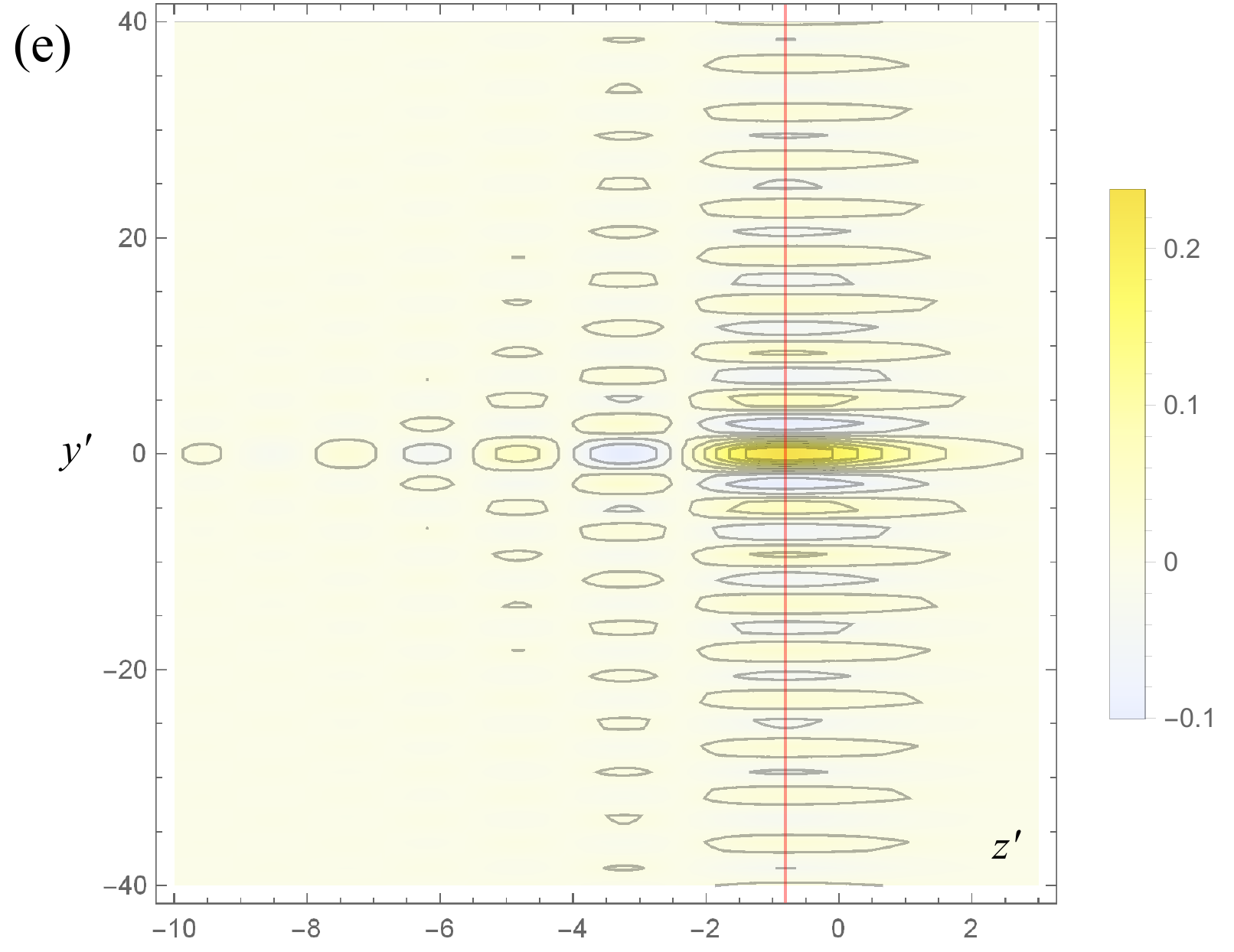} & \includegraphics[height=45mm]{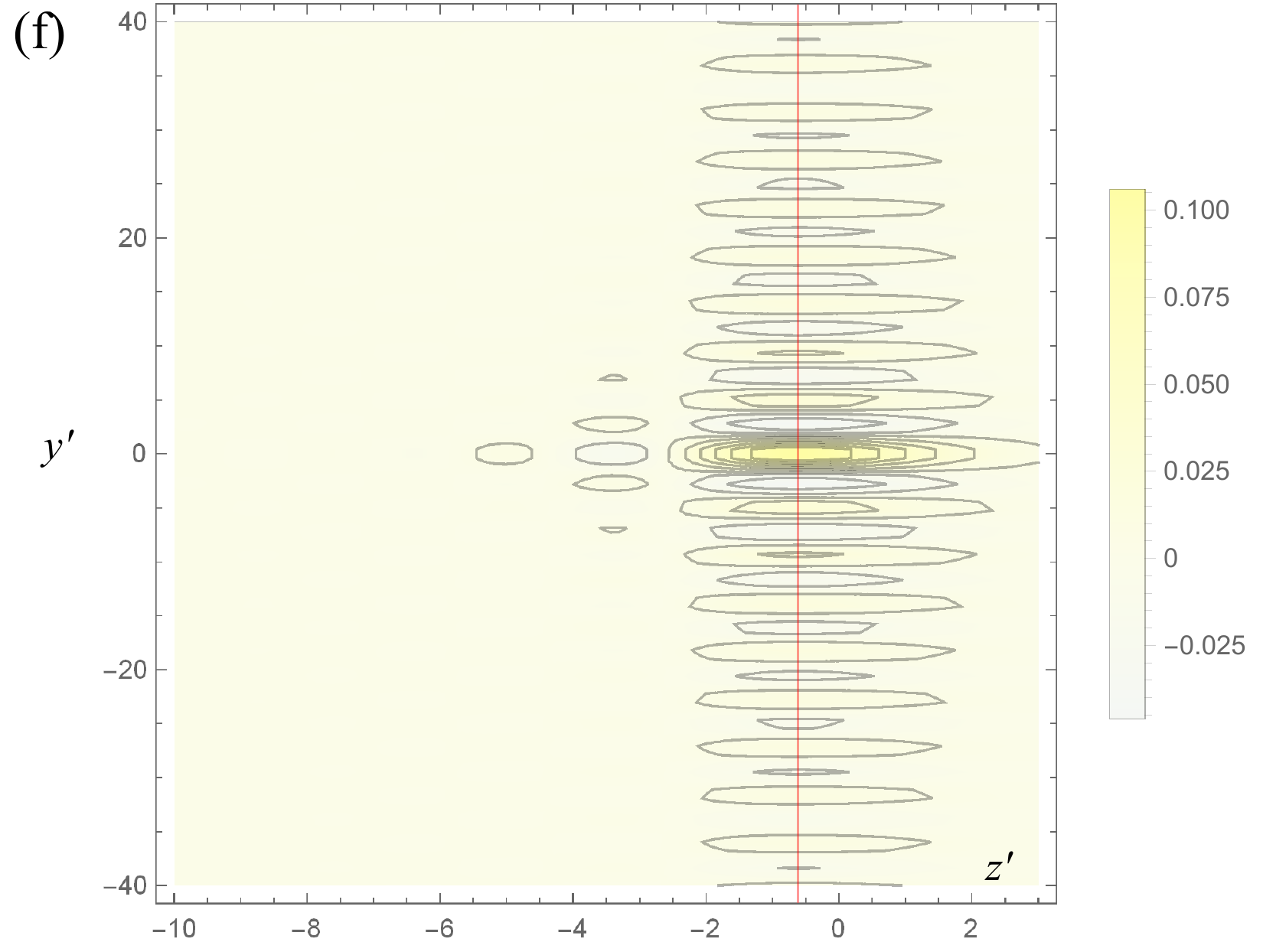}
\\
\end{tabular}
\caption{Density plots for diffusion densities \eqref{Airygauss} and  \eqref{soluDensiAiryBess}, showing the evolution of heat bullets solutions at three different times, in terms of variables $z'=k\, z/\sqrt{D_3}$ and  $y'=k\, y/\sqrt{D_2}$, in the plane $x=0$. Black curved lines represent the iso-contours for densities.
Red vertical lines represent the  position in $z$ for the maximum density lobe. (a) Airy-Gauss bullet   \eqref{Airygauss}, $\phi\sqrt{D_1D_2}/(\phi_0k^2)$, at
 $k^2 t=0.1$,  with maximum density lobe located in $z'=-0.926$. (b) Same Airy-Gauss bullet  at
 $k^2 t=0.15$,  with maximum density lobe  in $z'=-0.883$. (c) Same Airy-Gauss bullet  at
 $k^2 t=0.2$,  with maximum density lobe  in $z'=-0.842$. (d) Airy-Bessel bullet \eqref{soluDensiAiryBess}, $\phi/\phi_0$,  for $\lambda/k^2=2$ and at
 $k^2 t=0.05$,  with maximum density lobe located in $z'=-0.971$. (e) Same Airy-Bessel bullet  at
 $k^2 t=0.25$,  with maximum density lobe located in $z'=-0.804$. (f) Same Airy-Bessel bullet  at
 $k^2 t=0.55$,  with maximum density lobe located in $z'=-0.616$. }
\label{fig2}. 
\end{figure*}

The three-dimensional form of density \eqref{soluDensiAiryBess} is
shown in Fig.~\ref{fig1}(b) in the form  of an iso-contour plot, for  time $k^2 t=0.01$ and $\lambda/k^2=0.8$.
The acceleration of this Airy-Bessel  bullet is displayed in the form of density plots in Figs.~\ref{fig2}(d), (e) and (f). Here, the evolution of
solution  \eqref{soluDensiAiryBess} is shown for three subsequent times, in the plane $x=0$. We use vertical red lines to show the  motion of the maximum density lobe of Airy-Bessel solution.

Each maximum (and minimum) shows accelerating properties. In next section, we analyse such propagation for the maximum maximorum of this solution.

\section{Accelerating properties}
\label{accesection}

Both structured diffusion solutions,
\eqref{Airygauss} and 
\eqref{soluDensiAiryBess},
have accelerated motion. The
acceleration of the positions of maximum density on each lobe can be found by solving \cite{asenjohojmanT} \begin{equation}
    \frac{d {\mbox{Ai}}(\xi)}{d\xi}+k^2 t\,  {{\mbox{Ai}}(\xi)}=0\, ,
    \label{maxiEq}
\end{equation}
where
$\xi={k}z_M/{\sqrt{D_3}}+k^4 t^2$,
and $z_M=z_M(t)$ is the longitudinal time-dependent position of the maximum density on each lobe of interest.
We remark that the maximum density lobes
of both above  solutions fulfill Eq.~\eqref{maxiEq}.

\begin{figure}[ht]
  \includegraphics[height=50mm]{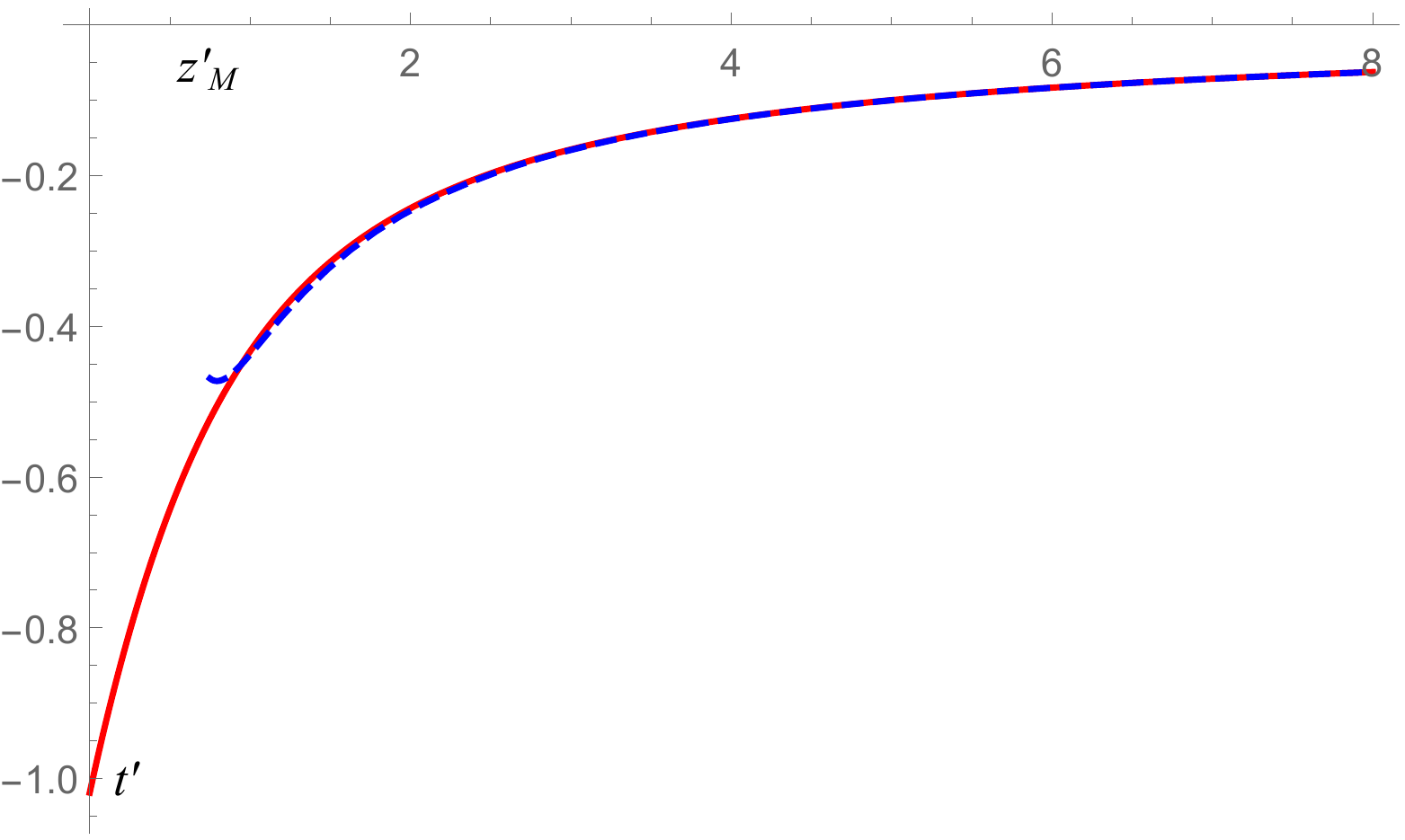}
  \caption{In red line, time-dependent position $z_M(t)$ of the maximum density of the main lobe of both solutions, \eqref{Airygauss} and \eqref{soluDensiAiryBess}, in terms of $z_M'=k z_M/\sqrt{D_3}$ and $t'=k^2 t$. In blue dashed line, analytical approximation \eqref{analyapprox} for large times.
}
\label{fig3}. 
\end{figure}

Let us focus in the position of the maximum maximorum density lobe. In this case, the accelerating behavior of $z_M$ is solved numerically, and it is shown in Fig.~\ref{fig3} in red solid line.
The plot shows how this time-dependent position $z_M$ evolves in time, showing that  $z_M\rightarrow 0$ when $t\rightarrow\infty$. It can be also deduced that the velocity of the maximum density
of the lobe (in terms of dimensionless units $k z/\sqrt{D_3}$ and $k^2 t$) is always positive but decreasing,
showing that the maximum  density of the lobe
decelerates.
 
 For long-times ($k^2 t\gtrsim 0.7$), the position of the maximum maximorum density of the main lobe can be analytically approximated by
\begin{eqnarray}
z_M\left(t\rightarrow\infty\right)\approx \frac{\sqrt{D_3}}{k}\left[\left(k^2 t-\frac{1}{4k^4 t^2} \right)^2 -k^4 t^2\right]\, .
\label{analyapprox}
\end{eqnarray}
This functionality is depicted 
as blue dashed line in  Fig.~\ref{fig3}. With the above expression we can obtain the acceleration for the maximum density of the lobe for longer times
\begin{eqnarray}
\frac{d^2 z_M\left(t\rightarrow\infty\right)}{dt^2}\approx -\frac{\sqrt{D_3}}{k^3 t^3}\, .
\label{analyapprox2}
\end{eqnarray}
Thus, the acceleration for maximum density (and therefore, for these  diffusive structured heat solutions) can be modulated by properly choosing $k$, for given $D_3$. It decreases as the position of the maximum maximorum density approaches $0$. 
A similar analysis can be performed for the other maxima of density solutions \eqref{Airygauss} and \eqref{soluDensiAiryBess}, which can be obtained by other solutions of Eq.~\eqref{maxiEq}.

Another simple diffusive propagating phenomena that can be exactly studied are the one for the voids, i.e., positions $z_v$ where the diffusive density vanishes. For the longitudinal motion, this occurs for the zeros of Airy functions in solutions 
\eqref{Airygauss} and \eqref{soluDensiAiryBess}. Different voids occur for different specific numerical values
of the arguments of Airy functions in both bullet solutions (such that  these values produce that the Airy function vanishes). The acceleration of those voids is analytically calculated to be constant and equals to
\begin{equation}
    \frac{d^2 z_v}{dt^2}=-2 k^3 \sqrt{D_3} \, ,
    \label{analyapprox3}
\end{equation}
for both solutions 
\eqref{Airygauss} and \eqref{soluDensiAiryBess}.
These previous accelerations \eqref{analyapprox2} and \eqref{analyapprox3}
show that localized parts of these diffusive bullets propagate differently. 

\section{Discussion}

The above diffusive accelerating solutions \eqref{Airygauss} and \eqref{soluDensiAiryBess} show analogue characteristics to light  bullets. This remarkable feature
emerges by the non-local properties introduced by Airy function as a solution of the linear diffusion equation \eqref{ecdifusion}. In this sense, the motion of the lobes are due only to the form of the solution, without any external source for diffusion 
or other different (linear or non--linear) modifications of the diffusion equation \cite{barcelona}.

These diffusive bullet solutions are determined by the initial conditions for the density $\phi(0,{\bf x})$. After that, the diffusion density evolves as it is shown above.
On the other hand, as the diffusion equation \eqref{ecdifusion} is linear, the above  diffusive solutions can be used to construct more complex bullet solutions with the form $\sum_i \phi_i$. Those may propagate in ways which are from the ones described here. 
Therefore, several diffusive accelerating properties of those bullets may be engineered at will.
On the other hand, we also expect that in more general equations, such as a hyperbolic heat diffusion equation for wave-like heat transport \cite{barcelona},  bullet solutions, similar to the ones discussed here, can be also found.

In conclusion, the accelerating localized diffusive solutions proposed here are new phenomena in the wider realm of diffusion, 
opening new directions on how heat can propagates. Consequently, it may be interesting to explore their impact in physics as well as in possible applications to medicine.


\end{document}